\documentclass[twocolumn,english,prc]{revtex4}
\usepackage[T1]{fontenc}
\usepackage[latin1]{inputenc}   
\usepackage{amsmath}
\usepackage{amssymb}
\newcommand{\beq}{\begin{equation}}
\newcommand{\eeq}{\end{equation}}
\makeatletter
\usepackage{graphicx}

\usepackage{babel} 
\makeatother
\begin{document}
\title{on the theory of supersolidity}
\date{December 12, 2007}
\author{Yongle Yu}
\email[email address:]{yongle.yu@wipm.ac.cn}

\affiliation{State Key Laboratory of Magnetic Resonance
  and Atomic and Molecular Physics, Wuhan Institute of Physics and Mathematics,
Chinese Academy of Sciences, Wuhan 430071, P. R. China}

\begin{abstract}
We present a microscopic, many-body argument for
 supersolidity. We also
illustrate the origin of rotons in a Bose system.

\end{abstract}
\maketitle
In physics the most fascinating phenomena
are perhaps
so-called macroscopic quantum phenomena, such as
superfluidity, partially due to that they
 are anti-intuitive 
at a visible scale.
It might look unlikely 
to get a clear theoretical picture of these 
phenomena 
since that quantum many-body problems 
are complex problems. 
However, some direct analysis still 
can be performed and
some understanding can be obtained at
 the very fundamental
level, for example,
the construction of
Laughlin's wavefunction for
fractional quantum Hall effect. In this paper,
we attempt to argue that supersolidity \cite{chan} 
is a natural
consequence of Bose exchange symmetry through 
a direct analysis of the many-body wave functions of
a Bose system, our analysis 
also reveals the origin of rotons
 in a Bose system, which is another
 fascinating issue in condensed matter physics. 
 
\begin{figure}
\begin{center}\includegraphics
{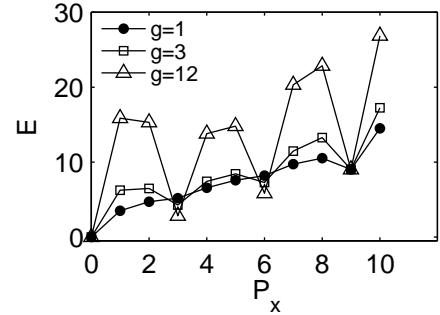}\end{center} \caption{The spectra of yrast states with
  $P_y=0$, relative to the ground state energy, for three systems with 
 $N=9$,  and having various interaction strengths $g$. the
 interaction takes a form of $f(\mbox{\boldmath$\theta$})=
 4 \pi^2 e^{-\theta^2/4\xi^2} /(4 \pi \xi^2) $ with $\xi= 2\pi/15$.}
\label{dispersion}
\end{figure}

The knowledge of superfluidity has accumulated through
the efforts of several generations of physicists \cite{tilley}. 
Landau \cite{landau} first
related superfluidity with the properties 
of the spectrum
of a system, specifically, the quasi-particle spectrum. 
Bloch's \cite{block} and Leggett's \cite{leggett}
  works clearly point out that  
a more transparent way to understand this phenomenon is
based on the properties of many-body dispersion spectrum,
 i.e., the lowest eigen energies of the system at given momenta.  
 When there exist local minima in the dispersion spectrum,
 the energy barriers which separate the minima will prevent
 the decay of current carried by the states corresponding 
 the minima (see Fig.~\ref{dispersion} for example), 
 thus superfluidity occurs \cite{largejump}.
 
\begin{figure}
\begin{center}
\includegraphics{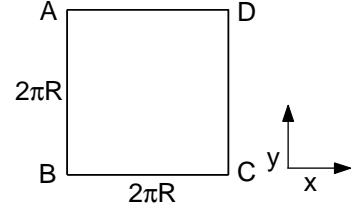}
\end{center}
\caption{ The geometry of a $2D$ periodic system.
 The edge $AB$ is identified with
$DC$ and $BC$ identified with $AD$.
 The linear size of the system is $2\pi R$.}
\label{2DGeo}
\end{figure}

 We have argued in \cite{yu} that the existence 
 of the local minima 
 in the dispersion spectrum of a Bose system  
 could be understood
 in terms of Bose exchange symmetry. Let us 
 consider
 a 2D periodic Bose system with $N$ particle, 
 with a finite-range 
 repulsive interaction and with a 
 square geometry (see Fig.~\ref{2DGeo}), 
 The Hamiltonian takes a form
 
 \begin{equation}
    H= - \sum_{i=1}^{N}  \frac{\hbar^2}{2MR^2}
     \frac{\partial^2}{\partial
  \mbox{\boldmath$\theta$}_i^2}  +  g \sum_{i<j}^N
 f(\mbox{\boldmath$\theta$}_i-
\mbox{\boldmath$\theta$}_j),
\label{Ham} 
\end{equation}

Where $M$ is the mass of a particle,
 $g$ is the interaction strength, $f$ describes
 the form of a repulsive interaction, $2\pi R$ is
  the linear size of the system. 
 $\mbox{\boldmath$\theta$}_i = (x_i, y_i)/R  $
 where $(x_i, y_i) $ are the coordinates 
  of the $i$-th particle. We will adopt the unit 
  system generated by $\hbar$, $R$ and $M$, except that
  the energy is in the unit of $\hbar^2/2MR^2$ where
  a factor of a half is involved for convenience.

The many-body spectrum can be labeled 
by the momentum
of the system ${\bf P}=(P_x, P_y)$, since $P_x, P_y$
are good quantum numbers due to the translational 
symmetry. We consider part of the spectrum, $E= E(P_x,
P_y)$, where $E$ is the minimum energy at the given 
$(P_x, P_y)$. We will call the many-body states 
corresponding this dispersion spectrum the yrast 
states, following convention in nuclear physics.

 Due to Galileo invariance \cite{block,yu}, we can
  derive the full yrast spectrum if 
  part of this spectrum
 in the regime $0\leq P_x \leq N$,
  $ 0 \leq P_y\leq N$ 
 is known. Particularly, we can derive all 
 local minima of the spectrum and
 realize that 
 there is an upper momentum limit
  beyond which
 no supercurrents exist any more, i.e., 
 there exists a critical velocity \cite{block,yu}.   
 We will only focus on the 
 structure
 of the yrast spectrum 
 in the momentum regime $0\leq P_x \leq N$,
  $ 0 \leq P_y\leq N$ in the rest of
  this paper.

The perturbation analysis of the yrast states
and dispersion spectrum of  the system 
at small $g$, is straightforward and presented 
in \cite{yu}. For example, the yrast states for 
$0 \leq P_x \leq N$, $P_y=0$ are 
 be approximated well by
Fock states 
$|(0,0)^N (1,0)^0\rangle,
 |(0,0)^{N-1} (1,0)^1\rangle, |(0,0)^{N-2}
(1,0)^2\rangle$, ..., $|(0,0)^0(1,0)^N\rangle$, 
where $(m,n)$, a pair of integers,
 denotes
the single particle orbit $\psi_{m,n}(x, y)=
 1/(2\pi) e^ {i m
x/ R + i n y/R}$ and 
$|(0,0)^{j_1}(1,0)^{j_2}\rangle$
denotes a Fock state with $j_1$ particles
 occupying
orbits $(0,0)$ and $j_2$ particles in orbits
$(1,0)$.  These Fock states will have the same 
(direct) Fock energy while the exchange 
energy, as a
function of $P_x$, has a parabolic behavior, 
$E_{ex} \propto  g (N-P_x)P_x $.  
This behavior of
the exchange energy, without any classical
analog, is responsible for the possible 
existence of local minimum of the 
dispersion spectrum
at $(P_x, P_y)=(N,0)$. In \cite{yu}, we also
 present the
second microsmany-body argument for the exchange
origin of superfluidity. By comparing the
dispersion spectrum of a spinless Bose
system and that of 
a corresponding spinor Bose system, one
can naturally imagine that
the possible energy barriers separating
two neighboring minima of dispersion spectrum
(of a spinless system) will become more apparent
with increasing $g$, i.e., strong interaction
enhances superfluidity.

In this paper, we will present another
many-body argument for the quantum origin of
superfluidity, which illustrate in a direct way
the role of exchange symmetry in determining
the dispersion spectrum of the systems. This 
argument also indicate a natural compatibility
of superfluidity and crystalline ordering, on 
the contrast to previous common beliefs that  
crystalline ordering will destroy superfluidity.
Supersolidity is a natural consequence of
this compatibility. 

We consider the system specified 
above with  $N=9$, the generalization for
$N=m^2$, where $m$ is an integer \cite{number}, is 
straightforward.
The dispersion spectra shows that with large $g$
there are also supercurrents
at $(P_x,P_y)= (3,0), (6,0)$ besides the one
at $(P_x,P_y)= (9,0)$ (see Fig.~\ref{dispersion}). The 
supercurrent states at  ${\bf P}= (3,0)$ and 
$(6,0)$ are rather surprising, in which it is 
impossible that all nine particles 
occupy the same orbits. This 'new' supercurrents
can be explained.

A many-body state of the system can 
be written in the spatial
presentation
\begin{equation}
|\psi \rangle = \sum_{x_1, x_2, ..., x_9}
         \psi(x_1,x_2,...,x_9) |x_1, x_2,..., x_9 \rangle
\end{equation}

one might refer to each state $|x_1, x_2,..., x_9 \rangle $
as a configuration with the first particle 
at position $x_1$,
the second at $x_2$, ..., the ninth at $x_9$.

and the interaction energy of the state has the form of, 
\begin{eqnarray}
E^\psi_{int}&=& \langle \psi |  g \sum_{i<j}^N
 f(x_i-x_j)|\psi\rangle \\
 &=& \sum_{x_1, ..., x_9}  |\psi(x_1, ...,x_9)|^2
  g (\sum_{i<j}^N
 f(x_i-x_j))
\end{eqnarray}                     
which can be 
interpreted as average of the interaction 
energy of each
configuration, with weight being 
the square module 
of the wavefunction. The configurations with
small interaction energy shall be largely 
explored by the yrast states 
at large $g$, in which the interaction
energy dominates the kinetic energy.

We shall consider a group of configurations 
with a certain discrete translational symmetry
and its relationship with yrast states. The
symmetry is following.
If one configuration is translated by
a mount of $2\pi/3R$ along $x$ axis, the 
first particles will be in the previous 
position of the second particle ($x_1 \rightarrow x_2$),
the second particles will reach the previous position
of the third particle  ($x_2 \rightarrow x_3$),
and  $x_3 \rightarrow x_1$,
$x_4 \rightarrow x_5 \rightarrow x_6 \rightarrow x_4 $,
$x_7 \rightarrow x_8 \rightarrow x_9 \rightarrow x_7 $.
We label the positions of nine particles of this 
configuration by $A_1, A_2, ...,A_9$, i.e., $x_1=A_1,
x_2=A_2, ..., x_9=A_9$. Thus after the translation, 
$x_1=A_2, x_2=A_3, x_3=A_1, ..., x_9=A_7$. 

\begin{figure}
\begin{center}\includegraphics
{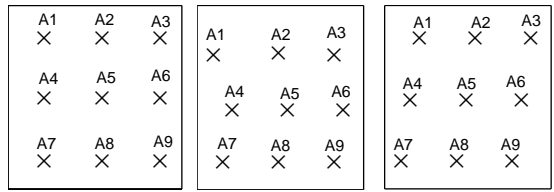}\end{center} \caption{
Some examples of configurations with
$2\pi/3-x-symmetry$ and with rather small
interaction energies. Crosses mark the positions denoted
by $A_{i}s$. }
\label{configs}
\end{figure}

We refer to this symmetry
as $2\pi/3-x-symmetry$. Obviously, most
configurations with $2\pi/3-x-symmetry$ has
a relatively small interaction energy since
the particles are well separated along $x$
axis, (see Fig.~\ref{configs} for examples). 

We shall argue that the configurations with
$2\pi/3-x-symmetry$ can only be accommodated
(with a non-zero weight) by yrast states at
 $P_x= 0,3, 6, 9,...$ .

First, with this translation, the
wavefunction associated with 
the configurations changes to be,
\begin{eqnarray} 
\Psi(x_1=A_2, ..., x_9=A_7) \equiv 
\psi(A_1 + b, ..., A_9+b) \nonumber \\
=e^{i b{\hat P}/\hbar}\psi(A_1,A_2,...,A_9)
\label{f1}
\end{eqnarray}
Where $b=(2\pi/3R, 0)$, $\hat{P}=(\hat{P}_x,\hat{P}_y)$
is the momentum operator. For eigenstates of 
the $\hat{P}$, $ b\hat{P}/\hbar= 
P_x/3*2\pi $ with $P_x$ in the unit we 
specified above.

Due to Bose exchange symmetry, the wavefunction
is invariant under any permutation of the
coordinates of particles, thus
 
\begin{equation} 
\psi(A_2, A_3, ..., A_7) =
\psi(A_1, A_2, ..., A_9).
\label{f2}
\end{equation}

Combining Eqs.~\ref{f1}, ~\ref{f2}, one is led to

\begin{equation}
(e^{i\frac{2\pi}{3} P_x} - 1)
 \psi(A_1,A_2,A_3;A_4,A_5,A_6;A_7,A_8,A_9)= 0.
\end{equation}

Thus $\psi(A_1, A_2, A_3, A_4, A_5, A_6,
 A_7, A_8, A_9)= 0$
at $P_x \neq 0, 3,6 ,...$ . The configurations
with $2\pi/3-x-symmetry$ can't be effectively 
involved in those yrast states with   
$P_x \neq 0, 3,6 ,...$ . 

One can also note that if one 
configuration is not involved 
in a yrast state $|\psi\rangle$,
its neighboring configurations can't be
effectively involved in  $|\psi\rangle$.
For a configuration with first particle 
at position $B_1$, the second 
particle in position $B_2$, ... , in its
neighboring configurations, the first particle is
at position $B_1+ \delta_1$ where $\delta_1$ is
small position shift, and the second particle
is at $B_2+ \delta_2$ with $\delta_2$ being small,
... . If $\psi(B_1, B_2, ..., B_9)= 0$, then
$|\psi(B_1+ \delta_1,B_2+\delta_2, ..., B_9 + \delta_9)|
$ can't be large, otherwise the value of the
wavefunction changes dramatically from
 $(x_1, x_2,...)=(B_1,B_2,...)$, to 
 $(x_1,x_2, ...)=(B_1+ \delta_1,B_2 + \delta_2, ...)$, 
 which
 means that the wavefunction has a large gradient 
 and thus a large kinetic energy (density) near this
 neighboring configuration, but large kinetic energy
 (density) shall be avoided in the yrast states
 which pursues the minimum energy at given momenta.

Therefore, the yrast states 
at $P_x=0,3,6,9,...$ can use the 
configurations with
$2\pi/3-x-symmetry$ and their
 neighboring configurations to 
 lower their interaction
 energy while the yrast states at
 other $P_x$ can not. Similarly,
 the yrast states at $P_y=0,3,6,9, ...$
 can effectively access configurations with
 $2\pi/3-y-symmetry$ and their 
 neighboring configurations 
 to lower the interaction
 energy while the states
  at other $P_y$ can not. At 
  sufficient large $g$, the yrast
  states at $(P_x,P_y)=(3,0),(6,0),(9,0),(3,3),...$,
  will become  supercurrent states, 
  with energies lower than the energies of their
  neighboring yrast states.
  
\begin{figure}
\begin{center}\includegraphics
{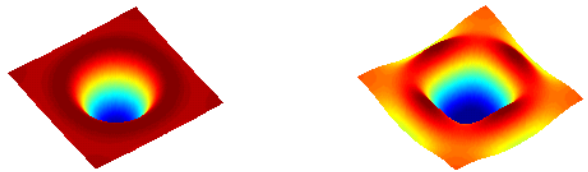}\end{center} \caption { Pair correlation profile of
 the ground state of a system with $g=1$ (left) and  of a system with 
 $g=12$ (right).
  The reference point
  $\mbox{\boldmath$\theta$}_A$
  is located at the
  center. $N=9$. At rather small $g$, $g=1$, the system 
  has short-range correlations but no long-range correlations and
  is in a gas-like or liquid-like phase (left). At large $g$,
  $g=12$, the system has long-range correlations and is
  in a solid-like phase (right). }
\label{paircor}
\end{figure}

  The configurations with $2\pi/3-x,y-symmetry$
  are also the fundamental components to
   build an inner crystal-like structure 
   in the system, one thus draw a conclusion
    that only those yrast states with 
    particular momentum 
    values can support such inner 
    crystal-like structures, 
   they are
   supercurrent states. 
   The inner crystal-like
    structure
    of the ground state at large $g$ 
    can be seen clearly
    in the plot of pair correlation function 
    (see Fig.~\ref{paircor}), which is defined as
    
\begin{equation}
    \rho (\mbox{\boldmath$\theta$}, \mbox{\boldmath$\theta$}_A)
     =  \frac {\langle \psi| 
  \sum_{i \neq j} \delta(\mbox{\boldmath$\theta$} - 
  \mbox{\boldmath$\theta$}_i) \delta(\mbox{\boldmath$\theta$}_A- 
  \mbox{\boldmath$\theta$}_{j}) |\psi\rangle} {\langle \psi | 
  \sum_{j} \delta(\mbox{\boldmath$\theta$}_A- 
  \mbox{\boldmath$\theta$}_{j}) |\psi\rangle}
\label{pair} 
\end{equation}       
where  $\mbox{\boldmath$\theta$}_A$ is 
the referees point.

\begin{figure}
\begin{center}\includegraphics
{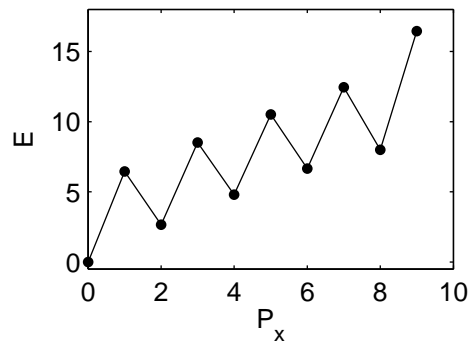}\end{center} \caption{ The yrast spectrum of  a 3D
  periodic system, relative to the ground state energy.   
$P_y=P_z=0$. $N=8$.  the interaction
 is $ V(\mbox{\boldmath$\theta$})=
 (2\pi)^3 g e^{-\theta^2/4\xi^2} /(\sqrt{4 \pi} \xi)^3 $, with $g=6$
, $\xi= 3\pi/20$, where 
 $ \mbox{\boldmath$\theta$} = (\theta_x, \theta_y, \theta_z)$
 is the difference between the generalized angular
  coordinates of two interacting particles. }

\label{3D}
\end{figure}

The above discussions can be generalized
 to systems with large $N$
and to $3D$  systems straightforwardly.
 In  a $2D$ system with square periodic boundary
  geometry, with $N= m^2$ \cite{number},
   and with large $g$,
   the yrast
   spectrum will has local minima
 at $(P_x, P_y)= (m, 0), (2m, 0)$,$...$, $(m,m)$, 
 $(2m,m)$,$
 ..., (m,m)$, due to the compatibility   
 of the yrast states at these momenta with
 the particle configurations with
  $2\pi/m-x,y-symmetry$.
 In a $3D$ system with cubic periodic boundary
 geometry and with $N=m^3$,  the yrast states
 at $(P_x, P_y,P_z)= (m, 0,0), 
 (2m,0, 0)$,$..., (m,m,0), (2m,m,0),
 ...$,$ (m,m,m)$, $...$, $(m^3, m^3, m^3)$ can become
 supercurrent states for their accommodations of
 the configurations with $2\pi/m-x,y,z-symmetry$.
 Numerical results of a $3D$ periodic cubic
  system with $N=8$ 
 is plotted in Fig.~\ref{3D} . 
  
 The above argument for exchange origin of
 superfluidity can also be applied
  to a
 system with an interparticle interaction which
 is repulsive at short range and attractive 
 at long range, for example, the Helium system. 
 In these systems, the configurations with that
 type of discrete translation symmetries are
 still energetically favorable and the states
 compatible with these configurations can become
 supercurrent states. One might naturally 
 speculate that supersolidity can 
 be observed in neon systems, much like
  the case of
 helium systems \cite{hydrogen}. 

\begin{figure}
\begin{center}\includegraphics
{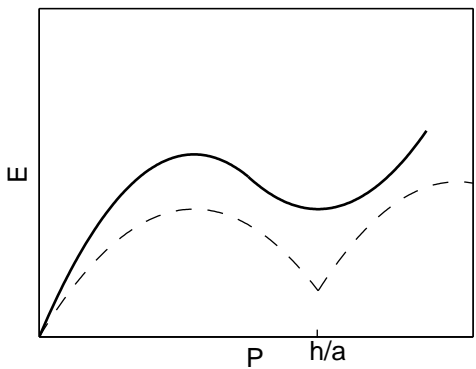}\end{center} \caption{Schematic plots of the 
quasiparticle dispersion (solid line) and yrast spectrum (dashed 
line) of a repulsive Bose system. Quasiparticles
generally do not exist beyond a threshold where
certain decay channels become allowed \cite{quasi-particle}. 
$h$ is
 Planck's constant.}
 \label{roton}
\end{figure} 
 
It is interesting to note that the smallest 
momentum of a supercurrent in
a  Bose system with large repulsive
interaction
 is $ \delta p=  N^{1/d}$ ($d$
is the dimensionality) and that it corresponds 
to the inverse of the average
particle distance $a$. In liquid $^4$He, the momentum
corresponding to  rotons in quasiparticle spectrum
is also around $h/a$ ($h$ is  Planck's constant),
 which naturally
 suggests that rotons are due to the local
minimum of the yrast spectrum. 
 The quasiparticle
 dispersion curve lies above the many-body dispersion
curve, as illustrated schematically in Fig.~\ref{roton}.
 Corresponding to the
local minima of the yrast spectrum at momentum $h/a$,
there are rotons in the quasiparticle spectrum.

 In conclusion, we present a microscopic many-body
 argument for supersolidity. We also predict 
 possible supersolidity in $^{20}$Ne 
  and in $^{22}$Ne.


\begin{thebibliography}{1}
\bibitem{number} One notes that the system can form an
        energetically favorable crystal 
        only at $N=m^2$. If the system  has different periodic 
        boundary geometry and/or different sizes, it will
        form a crystal at different $N$ values.  Since a crystal
        structure is expected at strong interaction regime, 
        we thus consider only the systems with these
         specific $N$ values.
\bibitem{largejump} Here one ignores
the probability for 
 a supercurrent state, a yrast state
  corresponding to a local minimum,  to decay into
  another supercurrent state (with a less energy).
In such a decay process, the system  emits a quantum
with a certain energy and with a large momentum
 which is at least of the order of
$h/a$ where $a$ is the average 
inter-particle distance. Depending on its own
low energy
spectrum, the environment may not be capable of
absorbing this quantum and 
thus this decay process is 
unlikely to occur.
\end{thebibliography}
 \end{document}